# ezBIDS: Guided standardization of neuroimaging data interoperable with major data archives and platforms

*Daniel Levitas[1,3*], Soichi Hayashi[1,3*], Sophia Vinci-Booher[1,2], Anibal Heinsfeld[3], Dheeraj Bhatia[3], Nicholas Lee[3], Anthony Galassi[4], Guiomar Niso[1,5], and Franco Pestilli[3]*

[1]Department of Psychological and Brain Sciences, Indiana University, Bloomington, IN 47408, USA

[2]Department of Psychology and Human Development, Peabody College, Vanderbilt University, Nashville, TN 37203, USA

[3]Department of Psychology, Department of Neuroscience, Center for Perceptual Systems, Center for Learning and Memory, Center for Aging Population Sciences, University of Texas, Austin, TX 78712, USA

[4]Center for Multimodal Neuroimaging, National Institute of Mental Health, Bethesda, MD, United States

[5]Instituto Cajal, CSIC, Madrid, Spain

**Corresponding author** Franco Pestilli (pestilli@utexas.edu)

**Acknowledgments**

This work was supported by NSF BCS 1734853, NSF BCS 1636893, NSF ACI 1916518, NSF IIS 1912270, and NIH NIBIB R01EB029272. Sophia Vinci-Booher was partially supported by a postdoctoral fellowship NSF SMA 2004877. Guiomar Niso was supported by the Spanish Government RYC2021-033763-I. We thank Chris Rorden, Hu Cheng, and Chris Markiewicz for responding to GitHub issues and emails pertaining to DICOM information, scanner parameters, and BIDS information, respectively. We also thank Kess Folco and Hu Cheng for helpful comments and suggestions. Lastly, we thank Elizabeth Berquist, Sandra Hanekamp, Kess Folco, John Purcell, Dan Kennedy, Taylor Zuidema, Megan Huibregtse, Nicole Keller, Frank Tong, Sam Ling, Taylor Salo, Dianne Patterson, Wade Weber, Giulia Berto, Damian Eke, Patrick Filma, Eberechi Wogu, Jie Song, and Heena Manglani for sharing data and testing early iterations of ezBIDS.

# Abstract

Data standardization has become one of the leading methods neuroimaging researchers rely on for data sharing and reproducibility. Data standardization promotes a common framework through which researchers can utilize others' data. Yet, as of today, formatting datasets that adhere to community best practices requires technical expertise involving coding and considerable knowledge of file formats and standards. We present ezBIDS, a tool for converting neuroimaging data and associated metadata to the Brain Imaging Data Structure (BIDS) standard. ezBIDS contains four major features: (1) No installation or programming requirements. (2) Handling of both imaging and task events data and metadata. (3) Automated inference and guidance for adherence to BIDS. (4) Multiple data management options: download BIDS data to local system, or transfer to OpenNeuro or to brainlife.io. In sum, ezBIDS requires neither coding proficiency nor knowledge of BIDS, and is the first BIDS tool to offer guided standardization, support for task events conversion, and interoperability with OpenNeuro and brainlife.io.

## Introduction

Data standards provide a framework for effective data sharing (Poldrack and Gorgolewski, 2014). Furthermore, data standards can help address replication issues in neuroimaging (Turner et al., 2018), by providing a common framework to harmonize data structures (Poldrack et al., 2017). A lack of data standards has downstream effects that can limit data sharing, curation, interoperability, and ultimately scientific reproducibility (Nichols et al., 2017). Without standards, attempts at data sharing create confusion due to an inability to readily discern the identity, location, and parameters of others' data.

The Brain Imaging Data Structure (BIDS) was introduced to provide a pathway to robust reproducibility and replication practices in neuroimaging by providing a widely accepted framework for standardizing the description and organization of brain imaging data (Gorgolewski et al., 2016). The BIDS standard has grown tremendously over the past several years and currently offers standardization specification for multiple neuroimaging modalities, including magnetic resonance imaging (Gorgolewski et al., 2016), magnetoencephalography (Niso et al., 2018), electroencephalogram (Pernet et al., 2019), intracranial electroencephalogram (Holdgraf et al., 2019), positron emission tomography (Norgaard et al., 2022), arterial spin labeling (Clement et al., 2022) and microscopy (Bourget et al., 2022) with many others under active development. The widespread adoption of BIDS can be attributed to the timely response to addressing data-sharing needs for scientific rigor and transparency (Gorgolewski et al., 2016; Reproducibility and Replicability in Science, 2019) and to the availability of community data repositories and platforms, such as OpenNeuro (Markiewicz et al., 2021) and brainlife.io (Avesani et al., 2019).

Notwithstanding the widespread adoption of BIDS, data standardization remains a challenge for many researchers, as it requires substantial knowledge of the BIDS specification and coding skills. Over the past few years, several software tools have been developed to assist researchers in the BIDS conversion process. These tools expect varying levels of manual intervention from the user, such as having users provide pieces of code or configuration files, including complex pattern matching statements, for improved BIDS mapping. A handful of tools promote limited manual intervention in the form of "point-and-click" graphical user interfaces (GUI), for example, a plugin for *Osirix/Horis*, the web-based *BIDS Toolbox (Lopez-Novoa et al., 2019)*, the *fw-heudiconv* package in Flywheel (Tapera et al., 2021), *pyBIDSconv*, *Biscuit* for MEG data, and *BIDScoin (Zwiers et al., 2021)*.

Despite the important features most BIDS-conversion tools provide, there remain both technical and knowledge hurdles that make converting neuroimaging data to BIDS a nontrivial task. Indeed, most BIDS conversion tools available require knowledge of the Unix/Linux command line (terminal) and moderate programming competency, for package installation and use. These requirements present technical barriers for researchers keen on adopting BIDS but who are more interested in specific domain questions and less inclined on learning new unfamiliar workflows. Furthermore, the limited built-in checks available in most BIDS-conversion tools to guide researchers to compliant BIDS datasets add uncertainty and limit adoption of the standard, with great loss for scientific rigor and transparency. To contribute to the adoption of the BIDS standard, it is necessary to develop conversion tools that support data conversions, while lowering the technical barriers of entry to BIDS by reducing the requirements for technical skills, such as coding and knowledge of the details of the standard specification.

In this article, we present ezBIDS, a BIDS-conversion tool that requires neither installation, programming skills, nor knowledge of the Unix terminal and BIDS specification. ezBIDS uses a Software-as-a-Service (SaaS) (Kiar et al., 2017) model where researchers can upload imaging data to a secure cloud system that hosts pre-installed software. The ezBDS web-interface guides users to organize entire sets of neuroimaging data series into BIDS-compliant datasets. Unlike other conversion tools that require users to input code or configuration files denoting the expected mapping (i.e., heuristics) of raw imaging data to BIDS, ezBIDS provides heuristics to the users. ezBIDS provides a guess for the mapping of the necessary information (e.g., *data type*, *suffix*, entity labels) for each image, and then displays the guess in a web-interface that allows users modifications and editing, if necessary. ezBIDS allows either to download data or push data to OpenNeuro or brainlife.io, contributing to lowering the barriers of entry to data standardization and sharing, while advancing scientific rigor and reproducibility.

## Results

ezBIDS facilitates BIDS conversion through the partnership between humans (user modifications via a web interface) and machines (heuristics to guess the mapping between neuroimaging data and the BIDS specification; **Figure 1**).

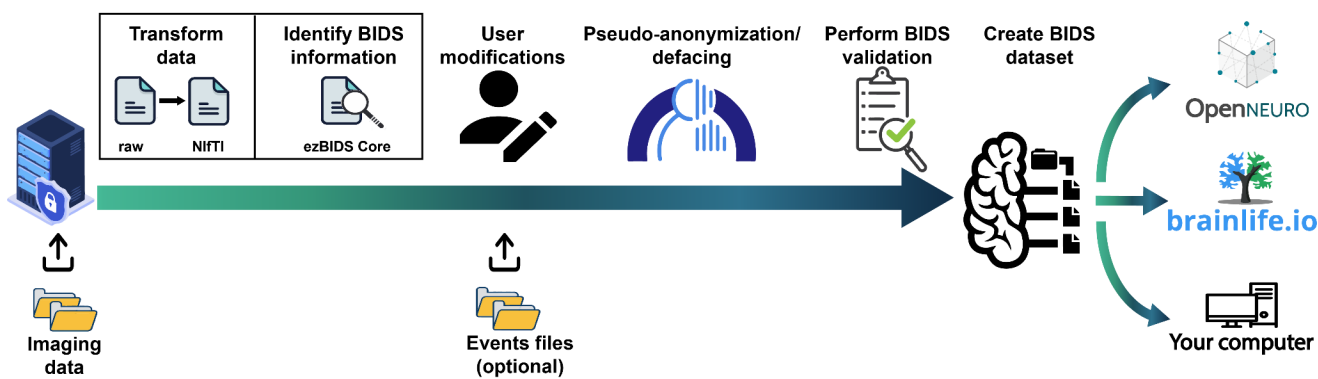

**Figure 1. ezBIDS workflow schematic.** The ezBIDS schematic overview of the steps necessary to map raw imaging data to BIDS. Users begin by uploading data to a secure ezBIDS server. Once uploaded, several automated backend processes transform the data and identify the relevant information required in BIDS specification. This information is then presented to the user for modifications, if needed. Following this, users may then choose to pseudo-anonymize anatomical data to remove identifying facial features (i.e., deface). ezBIDS then performs a final BIDS validation to ensure compliance with the specification, after which a finalized BIDS dataset is created. Finally, users may download their BIDS dataset to their local computer, or upload it to an open-science repository such as OpenNeuro, or to a data analysis platform like brainlife.io.

To ensure a standardized framework, the BIDS specification enforces specific rules dictating the organization and metadata necessary for a proper description of data. Firstly, BIDS requires that data be stored in specific formats. For example, MRI data must be stored in the Neuroimaging Information Technology Initiative file format (NIfTI; (Cox et al., 2004)) and the corresponding metadata must be stored using the JavaScript Object Notation (JSON) format (Crockford, 2006). NIfTI-formatted imaging data is compatible with a broad range of imaging software tools and smaller in size compared to other data formats (e.g., Digital Imaging Communication in Medicine (DICOM)) and JSON-formatted metadata provides relevant information that enables human and machine readability. Once formatted, files must be organized and named in a specific manner to enable the identification of the data. Files are placed in subject (and session, if applicable) folders, followed by *data type* (anatomical, functional, field map, diffusion-weighted imaging, etc.) sub-folders. Inside the *data type* sub-folders, files follow a naming convention that includes an entity labels list, a *suffix*, and a file extension. An entity is an attribute that describes the file, consisting of a hyphenated key-value pair, separated from other entities in the list by an underscore character. A *suffix* (commonly referred to as "modality"), denotes the sub-category of the *data type* (for example, *T1w*, *T2w*, and *FLAIR* are sub-categories of anatomical data). Lastly, the file extension denotes the data format, with *.nii.gz* for NIfTI and *.json* for JSON data. For example, a file named `sub-01/ses-01/anat/sub-01_ses-01_T1w.nii.gz` indicates that the data is an anatomical (*anat*) T1w sequence collected in the first session (`ses-01`) of the first subject (`sub-01`). Furthermore, the corresponding JSON sidecar metadata file (e.g., `sub-01/ses-01/anat/sub-01_ses-01_T1w.json`) contains data parameters such as the `SliceThickness`, `FlipAngle`, `ShimSetting`, among many others, which provide detailed information regarding its acquisition. In sum, BIDS provides a standardized framework for more effective data sharing and in support of scientific understanding and reproducibility.

The ezBIDS approach for mapping imaging data to BIDS is based on a Propose-and-Revise approach. The first step of this approach entails taking the input imaging data and automatically compiling BIDS-relevant information, which can then be mapped into a candidate BIDS structure (*Propose phase*). The information is gathered via an unsupervised process that uses a set of heuristic functions (see **Table S1**) implementing a traditional rule-based approach to produce an educated guess for the candidate BIDS-dataset structure given the set of input imaging files. The second phase of the approach is to allow revisions of the *Propose phase* by the user (*Revision phase*), which is presented to the user on the web interface. Drop-down menus and guided selections allow users to make revisions in accordance with the BIDS specification.

The ezBIDS workflow consists of several important steps that assist users in transforming raw imaging (e.g., DICOM) files to a BIDS-compliant dataset. This article reports the functionality of the ezBIDS system from both a user and a developer perspective. In brief, raw imaging data can be uploaded to a secure server hosted by brainlife.io, mapped to the BIDS structure, and then pushed to OpenNeuro, brainlife.io, or downloaded back to the user's computer (**Figure 1**).

## 1. Accessing ezBIDS

ezBIDS can be accessed at the URL https://brainlife.io/ezbids. DICOM files or the outputs of `dcm2niix` (Li et al., 2016) can be uploaded to ezBIDS for guided curation and standardization to BIDS by following the Web Interface at the ezBIDS URL. ezBIDS web services run on most browsers, but have been primarily tested and developed on Google Chrome, Apple Safari and Firefox. The process of using ezBIDS is demonstrated in this video: https://www.youtube.com/watch?v=mY3_bmt_e80. Below we describe the technical architecture of ezBIDS and the approach to data standardization and curation.

## 2. Upload raw data files: DICOM or `dcm2niix` output

ezBIDS has been developed and tested using two types of raw imaging data files: DICOM data as well as NIfTI and JSON data produced by `dcm2niix`. Data can be uploaded without any specific organizational structure, reducing user efforts for preparing data for conversion. Users may upload compressed imaging data using formats such as `.tar.xz, .tar, .tgz, .gz, .7z, .bz2, .zip,` and `.rar` (**Figure S1**).

Pseudo-anonymized DICOM data upload is supported, however, uploading DICOM data with sufficient unique information to allow mapping subjects and session IDs is preferred. Unique DICOM metadata allows reducing the burden on the users downstream during the conversion process. More specifically, if the metadata fields `AcquisitionDateTime`, `PatientName` and `PatientID` are preserved or at least containing unique information, they can be used by ezBIDS for mapping purposes. These fields typically provide unique subject (and session) information in imaging studies, which ezBIDS uses to organize the data with minimal user input. Note that these fields do not necessarily contain protected health information and it is up to the user to determine whether or not to anonymize the data prior to uploading to ezBIDS. ezBIDS can still process anonymized data lacking these participant-identifying fields but will likely exhibit less efficiency in determining the subject and session mappings, unless this information is explicitly contained (e.g., `sub-01`) within the file path. Explicit subject and session organization is not required for non-anonymized data, but providing the data organized in subject (and session) specific folders can help to regain some of the efficiency lost after anonymization. Uploaded data are private and non-accessible to other users and protected through the ezBIDS web-based security system (see **Web-security features of ezBIDS** section below). ezBIDS anonymizes the data once the BIDS conversion is complete, such that the data transferred to other cloud systems or downloaded to local computer environments are anonymized.

In lieu of DICOM data, users may instead choose to upload NIfTI and JSON data generated by `dcm2niix`. This option is convenient in cases where users have lost access to the original DICOM files. However, uploading DICOM data is generally recommended over `dcm2niix` data, as it contains more metadata facilitating BIDS information extraction, whereas for `dcm2niix` data there might be a mismatch between the current `dcm2niix` version used by ezBIDS and the potentially older `dcm2niix` version used by the user, which might not contain the most up-to-date metadata fields/information. In either case, ezBIDS accepts the data, performs a best-guess mapping to BIDS, and then guides the user through a series of web-based steps to assure that the mapping is accurate. Once uploaded, ezBIDS implements data uncompression if necessary, DICOM-to-NIfTI transformation using `dcm2niix` (if DICOM data are uploaded), and sends the output to the ezBIDS Core for BIDS (and additional) information identification.

## 3. The ezBIDS Core

The ezBIDS Core is a collection of python functions implementing operations to infer the mapping between raw and converted BIDS data files by parsing, interpreting, and organizing the data uploaded (**Figure 2**).

**a.** **Determine sub/ses IDs**
*determine_subj_ses_IDs*

**b.** **Group unique acquisitions by metadata**
*determine_unique_series*

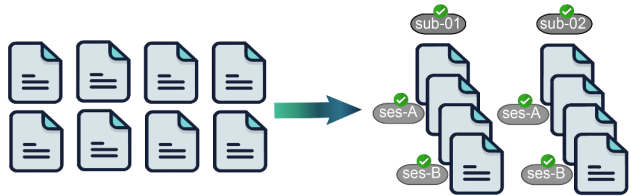

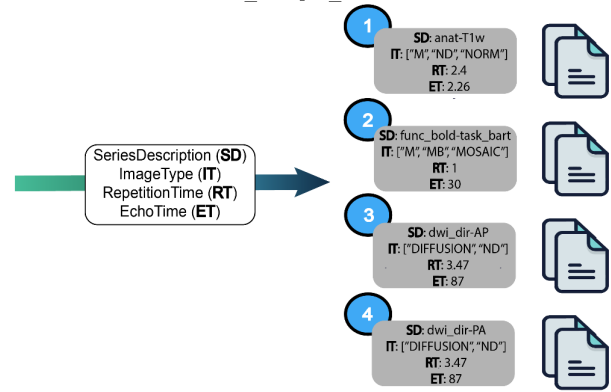

**c.** **Identify datatypes and suffixes**
*datatype_suffix_identification*

**d.** **Identify additional entity labels**
*entity_labels_identification*

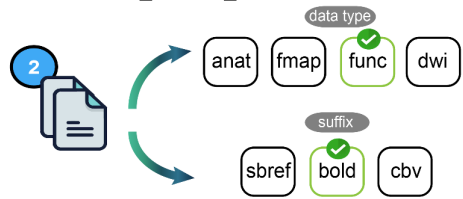

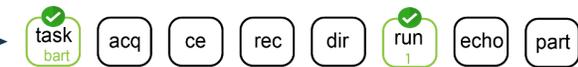

**Figure 2. The components of ezBIDS Core. a.** Function for determining subject (and session) BIDS entity labels. **b.** Function for organizing data into unique group series, based on having the same values for the following four metadata fields: `SeriesDescription`, `ImageType`, `RepetitionTime`, `EchoTime`. **c.** Function for determining the *data type* and *suffix* BIDS information, which provide the precise identity of the image. **d.** Function for determining additional BIDS information which provides a better understanding of the image's purpose.

ezBIDS Core gathers BIDS-relevant and additional information from the imaging data to present to the user via the web interface. ezBIDS Core is unique because it automatically guesses any available information contained within the imaging data provided, with the ultimate product being a JSON file called `ezBIDS_core.json`, containing metadata and inferred BIDS information (e.g., entity labels; **Figure 3**).

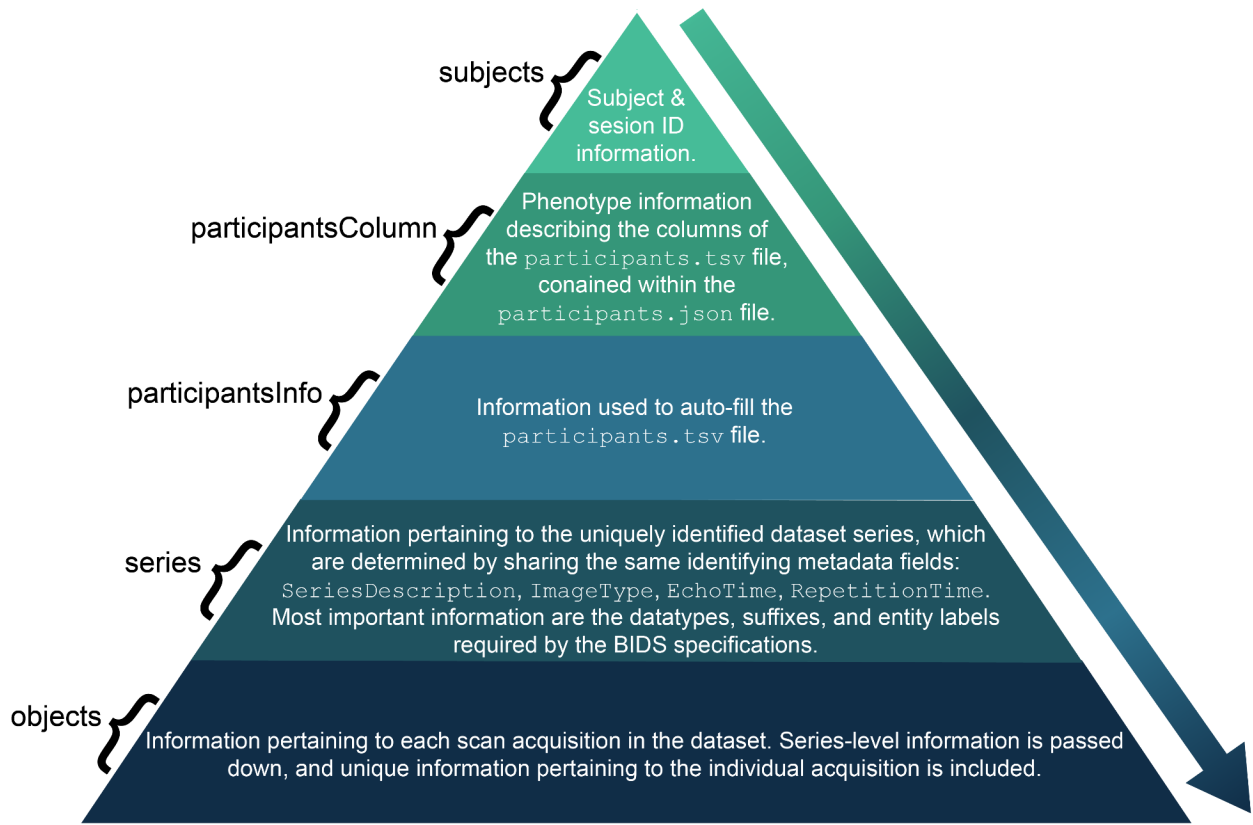

**Figure 3. Structure and contents of the `ezBIDS_core.json` file.** Schematic demonstrating the hierarchical structure of the JSON file created by *ezBIDS Core*. Each descending section comprises a larger proportion of the `ezBIDS_core.json` file. BIDS entity information (`sub-`, `ses-`, `task-`, etc) is passed down to levels of the individual scan sequences (objects). Objects'-level information constitutes the largest proportion of the JSON file, with subject information constituting the least. This JSON file represents an integrated representation of data and metadata mapping DICOM files to BIDS structures.

The `ezBIDS_core.json` is used by the ezBIDS web interface to display the BIDS information to the user, enabling edits, approval, and validation. This JSON file is automatically generated without requiring configuration files or predefined file naming conventions. The JSON file is output by the ezBIDS Core and can in principle also be used by other informatics systems as it contains BIDS information pertaining to each imaging file mapping, DICOM header information, data screenshots for visualization, and quality assurance information. Below we describe the steps that ezBIDS Core takes to generate the `ezBIDS_core.json` file.

(a) *Data transformation*. If DICOM data are uploaded, `dcm2niix` is executed, transforming the data to NIfTI format. Alternatively, NIfTI and JSON files generated by `dcm2niix` may be uploaded if the user does not have access to the DICOM files. While rare, any errors generated by the `dcm2niix` software are logged and presented to the user following this process, informing them which DICOM files(s) triggered an error and advising them to submit an issue on the `dcm2niix` GitHub page (https://github.com/rordenlab/dcm2niix/issues). Depending on the nature of the error, the offending DICOM(s) may not be transformed into NIfTI files for BIDS conversion. It is recommended that such errors be resolved by reaching out to the `dcm2niix` team; however, this is not required in order to proceed.

(b) *Generating BIDS information*. The challenge presented to all BIDS converters is to assign BIDS information to images, such as the data type, *suffix*, and entities (see **Table 1** for definitions of these terms). The ezBIDS Core is used to make a best-guess determination of this information, which is done through a series of functions (**Table 2**). There are four key functions that identify the appropriate BIDS information for each image (**Figure 2**).

| Term | Definition |
|---|---|
| *data type* | A functional group of different types of data. Data files are stored in a folder named for the specific *data type* (e.g., *anat*, *func*, *fmap*, etc). |
| *suffix* | The category of brain data recorded by a file. Often referred to as the "modality", it is a sub-categorization of the *data type* (e.g., *epi* is a *suffix* of the *fmap data type*). The suffix may overlap, but should not be confused with the *data type*. |
| *entity* | Label(s) that helps differentiate the data, based on image parameters (e.g., *acq*, *dir*, etc). Only specific entity labels are allowed for certain *data type* and *suffix* data, and the ordering of entity labels is necessary for BIDS compliance. |
| *series ID* | Unique value given to images based on specific metadata (`SeriesDescription`, `ImageType`, `RepetitionTime`, and `EchoTime`; see **Figure 2b**). |
| *task events timing files* | Files describing timing and other properties of events recorded during a run. For non-rest task functional BOLD images, BIDS highly recommends that such files be formatted as *events.tsv* files, containing standard information such as stimulus *onset*, *duration*, etc. |
| *validation errors (i.e., bids-validator errors)* | Situations in which an entry is not compliant with BIDS specification. Examples include improper entity label(s), misordering of entity labels, use of special characters, etc. Errors must be addressed to ensure BIDS-compliant data. |
| *validation warnings (additional ezBIDS-specific guidance messages and information)* | Situations in which some aspect of the data might be improper and worth addressing. Examples may include alerting users to exclude a sequence (e.g., *events.tsv*) from BIDS conversion because a corresponding sequence (functional BOLD) has been excluded. Warnings are meant to ensure improved quality of the dataset for curation purposes; however, they do not need to be addressed in order to create a BIDS-compliant dataset. |

**Table 1**: Definitions for BIDS-related terms. The terms *data type*, *suffix*, *entity*, and *task events timing files* come directly from the BIDS specification.

*b.1* Identify `subject` and `session` BIDS entity labels (ezBIDS *determine_subj_ses_IDs*; **Figure 2a**). Each image file name is required to have the `subject` entity, as well as the `session` entity when the participant data is spread across separate scan visits. ezBIDS examines each image's file path and name, using a regular expression (regex) search pattern to determine whether or not the `subject` and `session` entity labels are discernable from the file path. This determination is predicated on the existence of `<sub->/<sub_>` (for `subject`) and/or `<ses->/<ses_>` (for `session`) patterns, which are taken as explicit subject and session identifiers. If this determination cannot be made, image metadata typically contains several fields that pertain to the `subject` entity label. ezBIDS Core searches for the existence of these fields and if found, the precedence order for `subject` is `PatientName`, followed by `PatientID` and `PatientBirthDate`. *ezBIDS Core* searches for `<sub->/<sub_>` and `<ses->/<ses_>` patterns in these fields, and if none are found then the field's value is assumed to be the `subject` (and `session`) label. Detection of one or more sessions for a particular `subject` entity entails assessing the `AcquisitionDate` and `AcquisitionTime` fields. In accordance with BIDS, a `session` label is not generated in instances where a `subject` label contains a single set of `AcquisitionDate` and `AcquisitionTime` values unless the `<ses->/<ses_>` pattern is detected.

*b.2* <u>Identify and group similar images</u> (ezBIDS *determine_unique_series*; **Figure 2b**). Images are grouped together when they contain identical values for the following four metadata fields: `SeriesDescription`, `ImageType`, `RepetitionTime`, and `EchoTime`. ezBIDS Core assumes that images with these identical parameters are similar enough to require identical BIDS information, such as *data type*, *suffix*, or entity label(s). This grouping procedure enables a more streamlined process and is similar to how the recently published BIDS curation package (CuBIDS) accounts for the heterogeneity of parameters within datasets (Covitz et al., 2022). Each group of images is assigned a unique series ID, linking it to other images within the group. If ezBIDS or a user adds an entity label (e.g., `acq-mprage`; **Table 1**) to a single image, this modification is applied to all other images in the group. Later, users have the ability to make modifications to individual images, enabling more precise control.

There are two instances in which images can be grouped together while not containing the exact four metadata field values. The first case pertains to slight differences in the `SeriesDescription` values due to retroactive reconstruction of acquired images. Under circumstances in which normal image reconstruction fails, oftentimes due to insufficient scanner storage space, retroactive reconstruction is required to obtain the data in image space. The string `<_RR>` is typically appended to the end of the `SeriesDescription` field to denote this procedure, yet the image is functionally the same as other images containing the same `SeriesDescription` field. The second case pertains to minute divergences in `RepetitionTime` and `EchoTime` values, where a slight tolerance is given in instances where these values differ within a +/- 0.5ms (0.0005 sec) range, to account for slight precision differences in how compilers convert floating-point numbers from binary to ASCII. For example, an image with an `EchoTime` of 0.03 and another image with an `EchoTime` of 0.030001 are given the same unique series ID, assuming exact matches on the other metadata values. There are no instances in which grouping by the `ImageType` does not contain an exact metadata field.

*b.3* <u>Identify *data type* and *suffix*</u> BIDS information for images with unique series ID (ezBIDS *datatype_suffix_identification*; **Figure 2c**). A critical component for BIDS conversion is properly determining the *data type* and the *suffix* (also known as the "modality") of images, which pertains to the identity of the images (**Table 1**).

ezBIDS references the BIDS-specification schema, a collection of Yet Another Markup Language (YAML) files containing information regarding the data types accepted by BIDS and the *suffix* labels allowed for each specific *data type*. There exist three heuristics for determining this information. The first entails searching the image's `SeriesDescription` metadata field for the existence of an explicit *data type* and/or *suffix* identifier (e.g., *anat_T1w*). The second heuristic entails searching the image's `SeriesDescription` metadata field for the existence of common keyphrases used in MRI protocols (see **Table S2**). For example, "tfl3d" is a commonly used phrase to denote a *T1w* anatomical image, therefore the presence of this phrase provides sufficient information to discern both the *data type* (*anat*) and the *suffix* (*T1w*). The third heuristic entails using additional metadata fields, primarily *ImageType* and *EchoTime,* to discern *data type* and/or *suffix* identities. For example, the phrase "DIFFUSION" in `ImageType` indicates the presence of diffusion-weighted data (DWI), thus providing evidence regarding the *data type* (*dwi*) and *suffix* (*dwi*) of the image. Additionally, an anatomical image with an `EchoTime` exceeding 100 ms is assumed to be a *T2w*, thereby identifying the *suffix* (*T2w*). For a description of the metadata fields (`ImageType`, `EchoTime`, etc.) and their usage in ezBIDS, see **Table S3**.

ezBIDS begins with the first heuristic for determining the *data type* and *suffix* information, proceeding to the subsequent heuristics if this information cannot be discerned with the current heuristic. Should all three heuristic approaches fail to identify the *data type* and *suffix* for an image, its *data type* is set as `exclude` and left to the user to determine whether the data should be converted, helping to limit ezBIDS misidentifications that users may fail to notice. If the data are to be converted, the user will be asked to specify this information.

*b.4* Identify additional BIDS entity labels (via *entity_labels_identification*; **Figure 2d**). A BIDS entity label contains information regarding the data itself and/or its place within the hierarchical BIDS dataset structure. This feature enables users to quickly discern what kind of data is contained within the dataset and its location. Depending on an image's *data type* and *suffix*, additional entity label information may be required to differentiate images from one another. For example, BIDS requires the phase encoding direction ("dir") entity label for spin echo field maps. ezBIDS Core performs two passes to determine this information that is achieved through regex search patterns ("_<entityLabel>-") of `SeriesDescription` (see **Table S2** for examples). Searching other metadata fields can also produce information on the necessary entity labels. For example, `dcm2niix` produces an `EchoNumber` metadata field for multi-echo images; the ezBIDS Core looks for this field and if found, adds the integer to the `echo` entity label.

The overarching goal of the ezBIDS Core automated process is to gather necessary BIDS information and provide a first-level mapping between images and the BIDS layout. This information and mapping are encoded into the `ezBIDS_core.json`, ezBIDS parses the fields in the JSON file and presents the information to the user in a series of steps (i.e. web pages). These steps allow ezBIDS, in partnership with the user, to bring the BIDS-mapping process to completion.

## 4. Dataset Description

Once data have been uploaded and processed, the user interacts with the Dataset Description page. Alongside the `ezBIDS_core.json`, ezBIDS Core generates a BIDS

`dataset_description.json` file, whose contents are displayed at this page. BIDS specifies that the fields in `dataset_description.json` must include basic information such as the dataset name and version of the BIDS standard used for the conversion. By default, ezBIDS Core pre-fills the BIDS version field. Additional fields are optional but recommended, such as authorship and acknowledgments, funding information for the study, license agreements specifying how the data may be used by others, and proper citation for using the dataset. ezBIDS offers users the ability to provide this information through the web page to better describe their dataset for the neuroimaging community (**Figure S2**).

## 5. Subject and Sessions

Users then proceed to the Subjects and Sessions page, containing `subject` (and `session`, if applicable) entity labels. Modifications can be done manually or by choosing one of the options provided by ezBIDS through the "Reset Subject Mapping" button, consisting of three options: *PatientName*, *PatientID*, and *Numerical* (**Figure S3**). Selecting *PatientName* uses the `PatientName` metadata for the `subject` label. Selecting *PatientID* uses the `PatientID` metadata for the `subject` label, which may contain different information than `PatientName`, despite the similar attribute names. Selecting *Numerical* will create numeric values as `subject` labels, in chronological order. Numerical values less than 10 are zero-padded (i.e. 01, 02, etc).

## 6. Series Mapping

Data are organized by their unique series IDs, with BIDS information (*data type*, *suffix*, entity labels) detected from the ezBIDS Core displayed (**Figure S4**). Users may modify this information, which is then applied to all images with the same unique series ID. For example, a dataset containing 10 subjects where each subject contains a T1w anatomical image means that these 10 images have the same series ID. If a user adds "*mprage*" to the acquisition (`acq`) entity label of this series ID, then all 10 images with that unique series ID have the `acq-mprage` entity label applied. This is meant to alleviate time spent on modifications by the user; rather than having to apply the label individually to potentially dozens of images, the user only modifies one image with the unique series ID. Users have the ability to make changes to individual images upon reaching the Dataset Review page (see point **6** below).

An important aspect to note is how the `run` label is applied. If a dataset contains functional BOLD data, the images may all have the same unique series ID because they contain the same identifying metadata (*SeriesDescription, ImageType*, *RepetitionTime,* and *EchoTime*). Users are not required to set the `run` label at the Series Mapping stage, as ezBIDS automatically determines the `run` label and applies it for users to see on the Dataset Review page. It should be noted that the `run` label can be applied to all data types and suffixes, not merely functional BOLD images.

## 7. Events

To date, ezBIDS is the only BIDS converter that assists users in converting task events timing files from the experiment to the BIDS-specified *events.tsv* format that is necessary for analyzing task-based imaging data. The lack of this feature in other converters is likely due to the large variation in the formatting of timing files across researchers, labs, presentation softwares, and experimental paradigms. Although BIDS does not require task-based functional BOLD images to have corresponding `events.tsv` files, failure to do so results in task-based datasets that cannot be replicated. Given the unique structure of ezBIDS, which relies on a partnership between proposed conversions by machine learning algorithms and the revised conversions provided by the user, ezBIDS is able to alleviate conversion hurdles in task events timing files by providing an intuitive conversion mechanism that works well with a broad array of timing file formats (**Figure 4**).

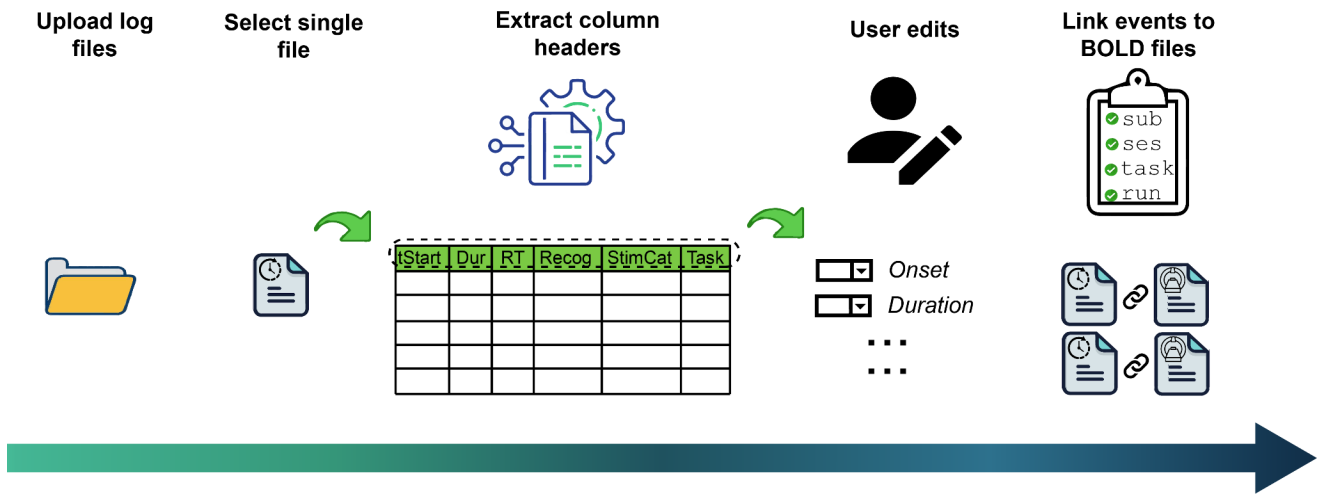

**Figure 4**. **Schematic of the ezBIDS events conversion process**. Once users have uploaded their task events timing files, ezBIDS performs a backend process to extract the column names based on the format of the uploaded files. The columns are then presented in a series of dropdown keys, enabling the user to specify which column name pertains to the BIDS task events columns. For time-based columns (e.g., `onset`), the user may specify "seconds" or "milliseconds" to note the unit of time for the recorded columns data. As BIDS requires time-based data to be in seconds, ezBIDS will convert data from millisecond to seconds, if so specified. Once the user has finished, ezBIDS applies the changes, converts the file format to TSV, and links the events files to the corresponding functional BOLD files by matching the `sub`, (`ses`, if applicable), `task`, and `run` entity labels. If ezBIDS cannot determine this link, unique random entity label values are provided (e.g., `sub-001`), which the user would then be able to edit. To enable greater accuracy in linkage, it is recommended that users specify these entity labels in the log file paths or as explicit column names.

Upon reaching the Events page, users may upload task events timing files, assuming they have task-based (i.e., non-rest) functional BOLD images (**Figure S5**). Accepted timing file extensions are `.csv`, `.tsv`, `.txt`, `.out`, and `.xlsx` and do not need to be formatted in any specific manner. Certain experiment presentation tools export their task events timing files in a specific format, such as E-Prime (Richard and Charbonneau, 2009), which can be handled by ezBIDS. Once uploaded, ezBIDS extracts the column names, which are presented alongside the BIDS `events.tsv` columns. Users then select which columns of their timing files pertain to the `events.tsv` columns; this action is performed once and then applied to all timing files, as it is assumed that all timing files for a scanning session will have the same formatting (**Figure S6**). Additionally, time-based `events.tsv` columns (*onset*, *duration*, etc.)

require values to be in seconds; however, some researchers output timing information in milliseconds. ezBIDS, therefore, provides the option for users to select whether or not their column values are in milliseconds and, ultimately, converts the values to seconds.

Once the crucial columns are identified, ezBIDS aims to link each timing file to its corresponding functional BOLD task image. This is done by looking for convergence between four BIDS entity labels: `sub`, (`ses`, if applicable), `task`, and `run`. ezBIDS assumes a one-to-one mapping of timing files to functional BOLD images, meaning that ezBIDS examines column headers or the timing file path name for information to identify appropriate entity labels for each timing file to map the timing file to the appropriate image (e.g., "Subject", "Sub", "Subj"). If a proper mapping to an image is determined, the timing file is given the same *SeriesNumber* as the corresponding BOLD data so that they will be displayed alongside each other on the ezBIDS web page. Mapping between timing and BOLD data can be rendered more efficient and require less user intervention by uploading timing files that contain explicit information about `subject`, (`session`, if necessary), `task`, and `run` either in their file paths (e.g., *sub-01/ses-01/task-bart_run-01*) or in their file columns .

Following at least one of these recommendations will result in the most efficient and automatic mapping possible. Should the mapping be imprecise – for example, a timing file cannot be automatically mapped to a specific subject – ezBIDS provides a placeholder value (e.g., `sub-XX1`), which the user then modifies to correct the mapping. Once modified, the web page is updated so that the *events.tsv* file is displayed alongside its corresponding functional BOLD image.

## 8. Dataset Review

At this stage of the ezBIDS conversion process all files are displayed by their `subject` (and `session`) entity label(s), and ordered chronologically (**Figure S7**). Unlike the Series Mapping page, where a single user modification affects all files in the group, here users make adjustments to specific files if need be. For example, a user may be aware that a specific subject's functional BOLD image is unusable due to excessive motion and do not wish to have the specific image converted to BIDS. The user has the ability to navigate to the data in question and exclude it from further BIDS conversion or add an entity label indicating the issue (e.g., `acq-highMotion`). For additional guidance, ezBIDS provides screenshots of each image and enhanced visualization examination through the `NiiVue` (https://github.com/niivue/niivue) plugin.

On the *Dataset Review* and *Series Mapping* pages, ezBIDS performs quality assurance checks to ensure adherence to the BIDS standard by executing the *bids-validator* and alerting users if some aspect of their data warrants closer inspection, specifically, validation errors and warnings (**Figure 5**).

**Figure 5. ezBIDS validation warnings and BIDS validator errors displayed on the WebUI.** ezBIDS provides validation errors from the BIDS validator as well as additional ezBIDS-specific warnings to the user. **a.** BIDS Validator errors are denoted in red and must be rectified by the user before proceeding. In the example, the user attempts to include a non-alphanumeric character ("."). into an entity label, which is not allowed by BIDS, hence, ezBIDS flags it as an error to alert the user to this inappropriate modification. **b.** ezBIDS warnings do not require user intervention to proceed but are meant to alert the user that intervention may be appropriate. In the example, functional BOLD acquisitions have been excluded from BIDS conversion; ezBIDS alerts the user that the corresponding SBRef and task events timing (*func/events*) files should also be excluded, otherwise their inclusion is unnecessary. Since such action is not necessarily required for a BIDS-compliant dataset, ezBIDS is agnostic with regard to how users respond to these warnings.

Validation errors are distinct from validation warnings. Errors are presented in situations where modifications or lack of information do not align with the BIDS specification, requiring correction before being allowed to progress. For example, entity labels cannot be alphanumeric; if a user attempts to specify the `acquisition` (`acq`) entity label with "0.8mm" then this is flagged by ezBIDS as an error. Users are unable to progress to the next page until all errors have been addressed. Warnings are generated in situations where the entity labels or data structure might not be proper; however, unlike validation errors, warnings do not require user intervention and may progress to the next page without addressing them. For example, if a user's uploaded data contains functional SBRef and functional BOLD images, and one of the BOLD images is excluded (i.e., set to `exclude`), ezBIDS will generate a warning for the SBRef file, suggesting that it be excluded as well because its corresponding BOLD image is also excluded. These validation warnings act as suggestions to improve the quality of the BIDS dataset, but users are not required to follow these recommendations.

## 9. Pseudo-anonymization/Defacing

Pseudo-anonymization is necessary for sharing data-sharing purposes (Eke et al., 2021). ezBIDS provides a web-interface to run software that can remove voxels containing face information from anatomical images (**Figure S8**). Two defacing options are supported by ezBIDS: *Quickshear (Schimke et al., 2011)* and *pydeface* (Gulban et al., 2019). *Quickshear* requires a skull-stripped brain mask as an input, which ezBIDS generates with the *ROBEX* tool (Iglesias et al., 2011). The recommended option is to use the *ROBEX + Quickshear* method on ezBIDS, as its performance is comparable to *pydeface*, but reaches completion in a noticeably shorter time. Before defacing begins, ezBIDS executes FSL's *reorient2std* function (Jenkinson et al., 2012) on each anatomical image to ensure that the defacing tools do not perform suboptimally due to uncommon image orientations (e.g., if an image were to be uploaded oriented Anterior-Left-Superior (ALS), it would be reoriented to Right-Posterior-Inferior (RPI) in order to match the MNI152 standard template orientation). While this procedure changes the anatomical orientations, this poses no issues for users, particularly ones interested in processing and analyzing their BIDS data with processing tools like *fMRIPrep* that automatically reorient anatomical images to the RAS+ orientation (Esteban et al., 2019). It should be noted that these defacing options are considered pseudo-anonymization steps due to advances in machine learning that could potentially result in re-identification of faces even from defaced anatomical images (Eke et al., 2021). Regardless, this should not deter users from performing defacing to better conceal participants' identity.

## 10. Participants Information

For each non-excluded subject, ezBIDS attempts to report basic phenotype information (age, sex) from the images' metadata (**Figure S9**). Users may populate these fields and/or provide additional participant information (e.g., handedness), which may be used as covariates in analysis models. Users may also choose not to include these fields and/or remove them.

## 11. BIDS dataset validation and download

All information generated by ezBIDS and modified by users is assembled to provide the data in the finalized BIDS directory structure and file naming convention (**Figure S10**). Users can choose to share their unique and private ezBIDS URL with collaborators to ensure that the data standardization or quality is satisfactory (e.g., verifying that the entity labels used are agreed upon by the research team). Lastly, ezBIDS reruns the *bids-validator* to ensure that the finalized dataset is BIDS-compliant, and displays the validator output to alert users to any lingering errors that may cause their data to be non-BIDS-compliant. Errors are displayed if the validator detects them, and while users are not prevented from creating a BIDS dataset, they are responsible for addressing the error(s). This is allowed because the error(s) may pertain to missing required metadata fields, which users cannot address through ezBIDS, thus the data is still converted, with the expectation that users will resolve the error(s) afterwards. If the user is satisfied with the final dataset, they have the option to download their BIDS-compliant dataset to their personal computer/server or upload the data to open repositories such as OpenNeuro (Markiewicz et al., 2021) or brainlife.io (Avesani et al., 2019).

## Web-security features of ezBIDS

As a web-based service that handles sensitive study participant imaging data, an important feature of ezBIDS is the secure implementation of its service stack. At the time of publication, ezBIDS runs on Jetstream2 Cloud, a HIPAA-aligned cloud computing infrastructure (Stewart et al., 2015). ezBIDS generates a unique key token for each new session that protects data from unwanted access. All ezBIDS sessions are anonymous, and stored data can only be accessed through an ezBIDS session URL with a unique key token. The communication between users and the ezBIDS web server is encrypted via `https` protocol using the *SHA256/TLS1.3* encryption algorithm. Uploaded data is temporarily stored on a dedicated *ceph* Jetstream2 volume, and all backend services are executed on a private VM responsible for the receiving, handling, processing, and downloading of imaging data through the *nginx* proxy. Jetstream2 provides a private subnet where the network traffic between the ezBIDS services can only be accessed by the member services. Users and collaborators with the proper URL have up to five days to work on their uploaded data, after which the data are purged from the storage system. Access to the private VM is restricted to only those who have the production ssh key as well as the administrators of the Jetstream2 Cloud who are trained IT professionals with proper HIPAA and system administration certifications. The activities within the VM are logged and monitored to detect and analyze improper use of the service.

## Comparison of ezBIDS to other conversion tools

ezBIDS seeks to broaden the adoption of BIDS to a greater audience within the neuroimaging community; however, there exist other BIDS conversion tools with the same stated goal. Like ezBIDS, these conversion tools have their own unique workflows and implementations for transforming raw imaging data into BIDS-compliant datasets. **Table 2** provides a comparison of ezBIDS to other commonly used conversion tools in the neuroimaging community along with a series of measures, though this is not an exhaustive list. Compared to other BIDS conversion tools, ezBIDS is the only tool with a web-interface, task events timing files conversion, QA checks, interoperability with OpenNeuro and brainlife.io, and a collaborative environment (via shareable ezBIDS session URL) with research team members that enables collegial BIDS conversion. ezBIDS is one of only two tools capable of embedding BIDS validation and being used without coding required from the user.

| | heudiconv | dcm2bids | bidsify | bidskit | Data2BIDS | BIDSCoin | ezBIDS |
|---|---|---|---|---|---|---|---|
| *CLI* | ✓ | ✓ | ✓ | ✓ | ✓ | ✓ | ✗ |
| *web interface* | ✗ | ✗ | ✗ | ✗ | ✗ | ✗ | ✓ |
| *Active development* | ✓ | ✓ | ✓ | ✓ | ✗ | ✓ | ✓ |
| *No coding required* | ✗ | ✗ | ✗ | ✗ | ✗ | ✓ | ✓ |
| *Task events timing files conversion* | ✗ | ✗ | ✗ | ✗ | ✗ | ✗ | ✓ |
| *QA checks* | ✗ | ✗ | ✗ | ✗ | ✗ | ✗ | ✓ |
| *BIDS validation* | ✗ | ✗ | ✗ | ✗ | ✗ | ✓ | ✓ |
| *Interoperability with open repositories* | ✗ | ✗ | ✗ | ✗ | ✗ | ✗ | ✓ |
| *Collaboration (shareable URLs) with colleagues* | ✗ | ✗ | ✗ | ✗ | ✗ | ✗ | ✓ |

**Table 2**: Comparison of ezBIDS to other common BIDS converters based on a collection of metrics.

## Discussion

Adoption of the BIDS standard promotes greater data sharing within the neuroimaging community by providing a common framework through which researchers can readily understand and utilize others' data. Data sharing is of great importance to the neuroimaging community for addressing the longstanding issue of underpowered studies and improving reproducibility (Button et al., 2013; Ferguson et al., 2014; Gorgolewski et al., 2017; Poldrack et al., 2017). Despite the clear benefits of data sharing (Milham, 2012; Poldrack, 2012) and a growing desire among researchers towards the implementation of FAIR principles (Kan et al., 2021; Poline et al., 2022), such practices are not yet ubiquitous throughout the neuroimaging community. This remains problematic, given the enormous financial costs of conducting neuroimaging research and the plethora of imaging data that already exists. One likely reason for the disparity between the practice and the aspirations of researchers are the high expectations necessary to practice FAIR data principles: Findability, Accessibility, Interoperability, and Reusability (Wilkinson et al., 2016). Specifically, data adhering to the FAIR principles must be described with rich metadata and be stored in a openly available resource/repository (Findable), must be available to others (Accessible), be compatible with resources such as softwares, repositories, databases (Interoperable), and contain accurate and relevant metadata containing clear and sanctioned data usage policies (Reusable). Such adherence adds an additional burden on researchers, who must already contend with the demands of their own research programs, funding sources, and teaching responsibilities, among others. The current work lowers the barrier of entry to practice FAIR principles by providing the neuroimaging community with a user-friendly and intuitive platform for converting unformatted file structures of raw imaging data into the BIDS standard ezBIDS.

A standardized framework for brain imaging data provides an ecosystem of supported-software tools that ingest BIDS data to perform automated, standardized processing and/or analyses, known as BIDS-apps (Gorgolewski et al., 2017). These tools aim to minimize the impact of variability in processing and analysis pipelines across labs and institutions, which can affect results and hinder reproducibility (Poldrack et al., 2017). A recent study exemplified this issue by demonstrating that processing and analyzing imaging data using three common yet separate software packages can lead to statistically different findings (Bowring et al., 2019; Poldrack et al., 2017). Furthermore, BIDS-apps provide standardized pipelines for processing/analyses that require minimal human intervention, due to the expectation of BIDS-compliant data as input. Access to BIDS data ensures that researchers can leverage BIDS-apps to reduce variability and save time in their processing and/or analysis pipelines.

ezBIDS lowers barriers for researchers who desire to practice FAIR data principles by offering many features that are not available in other BIDS conversion tools. Importantly, the unique features offered by ezBIDS target users who are more interested in working with a user interface for their BIDS conversion, instead of using scripting or a command line interface (CLI). Thus, a major strength of ezBIDS is that it does not rely on the CLI; however, it is possible that some researchers may consider this a limitation that hinders their ability to tailor ezBIDS to their specific needs. Future versions of ezBIDS may be developed further to include a CLI option for these users.

ezBIDS has been developed with MRI in mind, due to the authors' extensive experience with this imaging modality. However, the expansion of BIDS over the past several years now includes additional imaging modalities, including PET and M/EEG, among others. Future versions of ezBIDS will seek to include conversion capabilities for these additional imaging modalities.

As a web-based service, thorough testing has been performed on the Chrome and Firefox browsers; however, the performance of ezBIDS has not been extensively vetted on others such as Safari, Microsoft Edge, Internet Explorer, and Opera. Future work will conduct testing on these additional web browsers.

As data-sharing practices continue to gain widespread acceptance, it is critical to create tools that lower the barrier of entry to the BIDS ecosystem (e.g., BIDS conversion tools and apps). This entails offering conversion tools that do not solely rely on command line interfaces or specific programming proficiency, thereby reducing the burden of learning a particular converter's syntax and format. ezBIDS was designed with these considerations in mind, and as an innovative, open-source, web-based BIDS converter tool, aims to provide a streamlined and intuitive BIDS conversion experience to a broader range of neuroimaging researchers. As data sharing and adherence to the FAIR data principles become increasingly common, thanks in large part to BIDS, reproducibility issues in the neuroscience and psychological fields can begin to be addressed and will lay the foundation for new discoveries in the coming decades.

## Methods

The majority of the methods used by ezBIDS are described directly in the Results section above and in the supplementary material.

ezBIDS began as a series of custom scripts designed to semi-automate the process from DICOM data to BIDS-compliant data. Over time, additional back- and frontend processes were incorporated to relay information to users on a webpage, enabling point-and-click modifications as opposed to requiring programming usage from users. To validate the ezBIDS workflow, collaborators from academic institutions across North America and Europe provided sample imaging data from a variety of scanner manufacturers (Siemens, Phillips, GE), consisting of various data types: anatomical (`anat`), functional (`func`), field maps (`fmap`), diffusion weighted imaging (`dwi`), perfusion (`perf`), and scanner parameters. This provided a robust set of validation data to ensure that ezBIDS could accommodate a broad range of MRI data. Shortly after, beta-testers began using the service with collected data to find bugs and suggest enhancements for improving the ezBIDS experience. Developers would meet with beta-testers to discuss their experiences using ezBIDS to ensure that ezBIDS provided them with an intuitive user interface for BIDS conversion. Development was performed in an environment separate from the beta-instance used by the beta-testers to ensure that ongoing work does not interfere with users' interactions with ezBIDS.

ezBIDS has been validated using 30 shared datasets containing various types of data described in the BIDS specification from three scanner manufacturers. Errors and discrepancies in the mapping between DICOMs and BIDS structures for all these datasets and data types were used to inform updates on the ezBIDS code base and services.

# Data availability

Imaging data was privately shared and cannot be publicly disseminated. However, a public validation dataset was collected at Indiana University and is available at this Google Drive address: https://drive.google.com/drive/folders/1vVW3jPfYdvh52juiHTUnkmY6bjvxZQcR

# Code availability

All code is publicly available on our GitHub repository: https://github.com/brainlife/ezbids

To build and launch an ezBIDS development environment, `git clone` the GitHub repository, then execute `./dev.sh` in the root directory on a docker enabled machine. Once built and running, the dev instance can be reached at http://localhost:3000/. Alternatively, users can access ezBIDS by navigating to https://brainlife.io/ezbids/ in a web browser (Chrome or Firefox preferred), as there is no installation process.

### CRediT authorship contribution statement

**Daniel Levitas:** Conceptualization, Methodology, Software, Validation, Data curation, Writing–original draft, Writing-review & editing, Visualization. **Soichi Hayashi:** Conceptualization, Methodology, Software, Validation, Data curation. **Sophia Vinci-Booher:** Writing-review & editing. **Anibal Heinsfeld:** Software, Data curation, Validation, Writing-review & editing. **Guiomar Niso:** Validation, Writing-review & editing, Visualization. **Franco Pestilli:** Conceptualization, Methodology, Software, Validation, Writing–original draft, Writing-review & editing, Visualization, Supervision, Funding acquisition.

# Ethics declarations

**Competing interests** The authors declare no competing financial interests.